# Mapping Web Pages by Internet Protocol (IP) addresses: Analyzing Spatial and Temporal Characteristics of Web Search Engine Results

Ming-Hsiang Tsou*, Daniel Lusher*

* Department of Geography and Center for Human Dynamics in the Mobile Age (HDMA), San Diego State University, U.S.A.

**Abstract.** Internet Protocol (IP) addresses are frequently used as a method of locating web users by researchers in several different fields. However, there are competing reports concerning the accuracy of those locations, and little research has been done in manually comparing the IP geolocation databases and web page geographic information. This paper categorized web page from the Yahoo search engine into twelve categories, ranging from "Blog" and "News" to "Education" and "Governmental". Then we manually compared the mailing or street address of the web page's content creator with the geolocation results by the given IP address. We introduced a cartographic design method by creating kernel density maps for visualizing the information landscape of web pages associated with specific keywords.

## 1. Introduction

Web search engines are to some degree the most frequently used tools within the Internet. However, there is not very much published research characterizing these tools and their search results from a spatiotemporal perspective, despite widespread use in academic and professional studies. This research focuses mainly on the spatiotemporal patterns of search engine results, with a special emphasis on IP address geolocation accuracy. First, we look into how web search engine results are changing over time. While there is obvious temporal turnover in search engine results for the same keyword, we are interested in the rate of dropout for search engine results. Second, we ask about the relative broad categories of results returned by search engines. In a given set of search engine results, the percentages of blogs or government, or any other type of sites may vary. Next, we explore the spatial accuracy of the locations cited in IP address databases with respect to those of the content creators of search engine results. Finally, we introduced a cartographic method to create information landscapes of web pages, including setting kernel radius values, the method of calculating a population value, and the method used to normalize the data. By comparing the resulting maps side by side, the spatial patterns identified using IP addresses are better understood while recognizing better techniques to highlight the unique patterns associated with different keywords.

## 2. The Internet Search Engines, IP addresses, and Web Information Landscapes

One main idea in this research is that real-world information diffusion and events could be detected and visualized by online search engine results with associated keywords. While in the past, geographical information has been gleaned from methods such as surveys, field work, and remote sensing, this study utilized geo-referenced data from the Internet search engine results and their geolocations by referencing their Internet Protocol (IP) addresses. IP addresses is the unique identification number for each computing devices connected with the Internet. Some addresses are static, where it is the same over time and some addresses are dynamic, where the addresses change as devices log on or off the network. Each IP address is broken into bit lengths, with IPv4 (Internet Protocol version 4) addresses consisting of 32 bits (as is used in this research), and IPv6 (Internet Protocol version 6) addresses containing 128 bits. Many researchers, both in geography and outside the field, use IP addresses as a method for obtaining geolocational data within cyberspace.

Using a set of particular search keywords, disease outbreaks and events might be monitored quicker than standard government reports, such as Google Flu Trend (GFT) (Ginsberg et al, 2009). The validity of using internet searches to gather real world data has also been demonstrated in different sectors. The auto-industry, retail sales data, home sales, and personal travelling patterns can all be predicted faster than traditional reports by using search engine data (Varian and Choi, 2009). By expanding these basic trends into an entire geospatial cyberspace infrastructure, as discussed later, these patterns can become even more pronounced and identified.

Validity is a key issue when embarking on a research project that is founded upon a method that has not been verified by an outside source, and has not been fully exposed to the greater scientific community. While search engine data has shown to be useful in the examples above, is there any reason to depend upon IP addresses for locating web servers? In the absence of a non-automated research study on this topic, this research includes a component of manual classification of IP addresses, comparing the IP address location to the web site stated location. There has been some research into the accuracy of IP address geolocations, especially into the accuracy of databases such as MaxMind (which is used in this study). Shavitt and Zilberman (2010) take an exhaustive look at several different IP address databases, and attempt to examine the accuracy of each. In order to assess, they use 'PoPs', or points of presence, as the ground truth for IP addresses, and check to see when multiple IP addresses are geolocated to the same PoP for each database. Their results for MaxMind show that the database tends to have a NULL classification for 36% of the IP addresses, which includes both situations where no results are returned and situations where MaxMind returns the coordinates of the country's center (Shavitt and Zilberman, 2010, p.7). When looking at the MaxMind accuracy to a point, the MaxMind database has a "probability of 62% to 73% to place an IP within 40km from the PoP majority vote, with IPligence and MaxMind placing over 80% of the IP addresses within a 500km radius." (Shavitt and Zilberman, 2010, p.9). This indicates that one should expect the geolocation accuracy of IP addresses to be roughly within a city 62% to 73% of the time.

The search engine is the primary vehicle for pinpointing web pages in this research. The best method for search engines to generate results is PageRank. PageRank is the method used by Google to rank the relevance of the search engine results given a particular search. By objectively ranking a page based on popularity, not on the subjective content, PageRank can replicate the probable needs of a user, along with the tendencies of a random web surfer. This 'popularity' is determined by hyperlinks, both hyperlinks pointing to a particular page, and hyperlinks by a page pointing to others (Page et al, 1998).

Chen et al (1996) discuss a way to improve search engine performance by cataloging certain concepts before running the search. Since the versions of search engines in use in 1996 were very slow at examining each web page, Chen et al looked to build a system that would first categorize each homepage for a website, and then let the search engine search the categories. We borrowed the concept from Chen's research to classify websites into twelve categories (such as blog, news, government, etc.). Within each category, the IP address accuracy and the overall spatial patterns in the visualizations are compared. This categorization process is related to the idea from Chen et al in that this research involves trying to break a large topic into smaller pieces based on the semantic clues from the data.

Another related field of this research is geographic information retrieval (GIR) (Purves et al. 2007; Jones and Purves 2009). The scope of geographic information retrieval ranges from the detection of geographic content on the Web (Markowetz et al. 2005) to the analysis of IP geolocations (Buyukokkten et al. 1999), to the search engines of geotags (Amitay et al. 2004) and gazetteer reasoning databases (Silva et al. 2006). A seminal work in GIR, Purves et al. (2007) "*The Design and Implementation of SPIRIT: a Spatially-Aware Search Engine for Information Retrieval on the Internet*" introduced the design, implementation, and evaluation of a spatially aware search engine. The prototype identified geographic references from web pages (documents) and automatically created spatial footprints to index the contents. By using web crawlers (Joho and Sanderson 2004), geographical ontology databases, gazetteer lookup services, and geoparsing engines, SPIRIT can index and rank web documents based on their textual and spatial relevance (Purves et al. 2007).

Privacy is also one major concern in several Internet applications. Generally, people see their activity online as semi-anonymous, given that they are not doing anything infamous and there are so many other people online. However, user locations can be gathered fairly easily via IP addresses, as is used in many web applications. Issues surrounding privacy within web surfing need to be considered when approaching a proposal regarding mapping the locations of web servers (Svantesson, 2005).

## 3. Web Page Data Collection: IP Geolocation Methods and the CyberDiscovery Tool

IP geolocation (Muir and van Oorschot 2009) is a popular technique for identifying geographic location of Internet users or web servers. Researchers can convert IP addresses into real world coordinates (latitudes and longitudes) or geographic regions by using IP geolocation methods. The geolocation analysis of website visitors has become an important component in Web log analysis research (Fleishman 1996; Turner 2004; Backstrom et al 2008) and has been applied in various domains, including Location-Based Service (LBS), target marketing, epidemiology, and criminal investigation (Choi and Tekinay 2003; Lee 2008; Tsou and Kim 2010). IP geolocation is a cited method for implementing a framework based on Knowledge Discovery in Cyberspace, a framework which combines place, time, and messages (Tsou and Leitner, 2013).

There are two types of IP geolocation techniques: active IP geolocation and passive IP geolocation. Active IP geolocation technique relies on the om one IP address to another. However, the active method requires complicated calculations and cannot handle a very large volume of IP geolocations. Passive IP location is a database-driven procedure which relies on relational databases (such as MS SQL or MySQL databases). The IP geolocation databases include the index for mapping different levels of IP address (blocks or prefixes) to countries, cities, zip codes, and real world coordinates (Poese et al. 2011). For example, the database can convert the IP address, 130.191.118.3 to the U.S. zip code: 92182. The database also includes the latitude and longitude coordinates of the central point of zip code polygons.

| Rank | Search Engine | Keyword | Search Date | URL | Type |
|---|---|---|---|---|---|
| 4 | Yahoo | Jerry Sanders | 09/08/2011 | http://www.sandiego.gov/mayor/about/ | govermental |
| 5 | Yahoo | Jerry Sanders | 09/08/2011 | http://features.rr.com/topic/Jerry_Sanders | News |
| 6 | Yahoo | Jerry Sanders | 09/08/2011 | http://www.daylife.com/topic/Jerry_Sander | blog |
| 7 | Yahoo | Jerry Sanders | 09/08/2011 | http://www.huffingtonpost.com/jerry-sande | News |
| 8 | Yahoo | Jerry Sanders | 09/08/2011 | http://www.zimbio.com/Jerry+Sanders | News |
| 9 | Yahoo | Jerry Sanders | 09/08/2011 | http://topics.treehugger.com/topic/Jerry_Sa | Special Interest |
| 10 | Yahoo | Jerry Sanders | 09/08/2011 | http://www.amdboard.com/sanderspecial.h | commercial |
| 11 | Yahoo | Jerry Sanders | 09/08/2011 | http://twitter.com/MayorSanders | Social Media |

| Type | Latitude | Longitude | ZipCode | Area | IP | Host |
|---|---|---|---|---|---|---|
| govermental | 32.7977 | -117.23 | 92109 | 858 | 198.180.31.12 | sandiego.gov |
| News | 40.7619 | -73.976 | | 212 | 64.147.115.89 | features.rr.com |
| blog | 40.7619 | -73.976 | | 212 | 64.147.115.80 | daylife.com |
| News | 40.6888 | -74.02 | 10004 | 212 | 69.60.14.6 | huffingtonpost.c |
| News | 29.5072 | -98.575 | 78229 | 210 | 67.192.149.194 | zimbio.com |
| Special Interest | 40.7619 | -73.976 | | 212 | 64.147.115.89 | topics.treehugg |
| commercial | 37.4249 | -122.01 | 94089 | 408 | 98.139.135.21 | amdboard.com |
| Social Media | 0 | 0 | | | 199.59.149.198 | twitter.com |

**Figure 1.** The search results of "Jerry Sanders" keyword from the CyberDiscovery Tool. Note: the "Type" field is coded manually by graduate students in this study.

Our research team created a new tool, called "CyberDiscovery Tool" by combining multiple search engine's APIs (Yahoo and Bing) (Tsou et. al. 2013). Using the Yahoo and Bing APIs, a python script takes the user-inputted keyword to search for the corresponding ranked web pages. The top several hundred results are collected automatically into a MySQL database. The tool gathered a variety of web page information, including URL, IP Address, and Host Name (Figure 1). The raw list is geocoded using another python script that utilizes the MaxMind free geocoding database to match IP Addresses to the real-world location. MaxMind is a company that specializes in IP address geolocation and has developed a very large database of IP addresses with the corresponding server-registered geographic addresses. The final result of the CyberDiscovery tool is an Excel spreadsheet with web page data, locational data, and ranking data. This Excel spreadsheet can then be processed in a GIS. Using ESRI's ArcMap, part of the ArcGIS software package, the spreadsheet data is plotted using the coordinate information. Figure 1 illustrated an example of the keyword search results by using the CyberDiscovery Tool with keyword "Jerry Sanders" (Jerry Sanders was the mayor of San Diego in 2011).

Tsou, M. H., & Lusher, D. (2015). Mapping Web Pages by Internet Protocol (IP) addresses: Analyzing Spatial and Temporal Characteristics of Web Search Engine Results. *Proceedings of the International Symposium on Cartography in Internet and Ubiquitous Environments* 2015, 17th - 19th March, The University of Tokyo, Japan.

## 4. Temporal Analysis of Web Search Results

To examine the temporal differences within search engine results, we have compared the contents of a particular set of search engine results with the search engine results from the same keyword after a few days or weeks. In particular, we were interested in the decay rates of search engine results. In order to test this, we used the CyberDiscovery tool to search a keyword, and then search that keyword approximately each week for several months. Once we had these search results, we used a series of joins (table connections based on a key field) to find the number of results with the same URL in each week of results. We can show the decay curves in terms of identical URLs from week to week (Figure 2) and the number of results that have the same search engine rank from week to week.

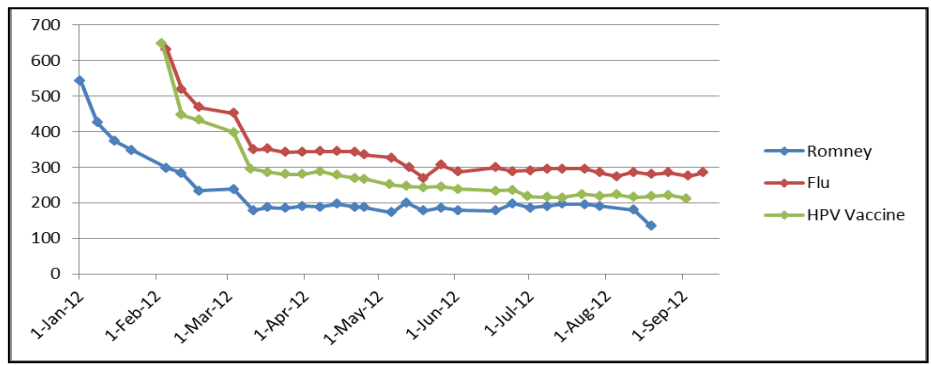

**Figure 2.** The change of web page search results for diffreent keywords (Romeny, flu, and HPV vaccine) in Yahoo Search Engine. The number indicates the number of same URLs comparing to the first week.

The keywords used for the temporal analysis are below:

- "Mitt Romney" – Thirty weeks search results ranging from January 1, 2012 through August 19, 2012.
- "Flu" – Thirty weeks search results ranging from February 5, 2012 through September 9, 2012.
- "HPV Vaccine" – Twenty-nine weeks search results ranging from February 3, 2012 through September 3, 2012
- "Osama bin Laden" – Twenty-two weeks search results ranging from May 4, 2011 through September 10, 2011
- "burn Koran" – Eight weeks search results ranging from April 3, 2011 through August 2, 2011

When a user types a set of keywords into a commercial search engine, the results are the product of the search engine at that moment. We compared the search results for a particular keyword over time, approximately on a weekly basis, and then looked at the number of records that had the same URLs (web pages) comparing to the first week search results.

There is a relatively quick dropoff in identical URLs (the same web pages) during the first few weeks as we expected. However, there is also an equilibrium that each keyword eventually hits (after 6 to 10 weeks). The level of the equilibrium may also be dependent on the dynamism of the keyword used. Dynamism would refer to the rate of interest in a keyword, with a more dynamic keyword leaving the news cycle faster. More dynamic keywords will have a lower number of same results. The Romney example above reflects this, as his campaign was in full gear throughout this time period. Figure 3 illustrates a temporal change comparison of two dynamic keywords ("Osama bin Laden" and "burn Koran") with Yahoo Search Results.

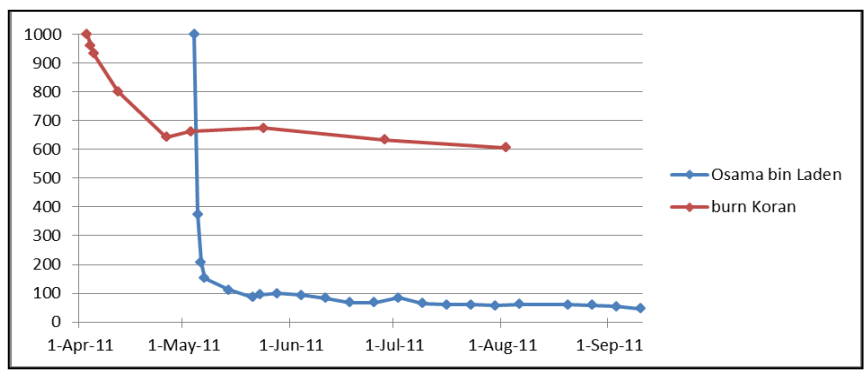

**Figure 3:** A temporal change comparison of the web search results with two dynamic keywords ("Osama bin Laden" and "burn Koran") with Yahoo Search Results. The number indicates the number of same URLs comparing to the first week search results.

The searches for "Osama bin Laden" began immediately one day after his death (2-May, 2011), which was a time of high interest for this keyword. This high dynamism results in a lower equilibrium, while the less newsworthy term "burn Koran" has a higher equilibrium level. This sharp drop-off followed by a plateau is shown in many of our examples.

We also examined the number of records within the search engine results that had the **same ranking** within the results as the first set of results. There are very few records that share the same rank as the original set of search engine results. There are generally a higher number of identical rankings in the second set, and then the count of identical ranks stays under ten for the rest of the sets of results. There does not appear to be much information to be gained from examining the records with the same rank.

## 5. Manual Classification of Web Search Results

Due to the limited human resources, we can only select six sets of keyword search results to perform the manual classification. We selected one representive week for each keyword. Web search results from each keyword contain over 600 to 1000 web pages per search result. We selected the following six keywords for this manual classification process (around 600 x 6 = 3,600 web pages):

- "Mitt Romney" – Search performed on January 4, 2012
- "Rick Santorum" – Search performed on January 4, 2012
- "Michael McGinn"- Search performed on November 17, 2011
- "Jerry Sanders" – Search performed on September 8, 2011
- "Flu" – Search performed on May 6, 2012
- "HPV Vaccine" – Search performed on February 3, 2012

After the web search results have been gathered, each result was processed using our Manual Classification Process. For each web page, three different data points were gathered. First, the web page was categorized into one of twelve **categories**. By placing the web pages into certain categories, the web pages can be scrutinized to see if there are any patterns based on web page type. The 12 categories of web pages are:

1. Blog (blog + personal + group): personal blogs or group blogs, usually 'Blog' has clear index about who are the writers and when the article was posted, along with personal web pages with strong opinions. 'Blog' is also applied when it was a site that was produced by a popular blogging company- such as WordPress or Blogspot.
2. Commercial websites: (e.g., selling products or selling services, or explaining information related to commercial products).
3. Educational: e.g., schools, universities, and educational institutes, usually within the EDU domain.
4. Entertainment + video: 'Entertainment' is used when there is only a video or series of videos available, not attached to an article, such as Youtube.com. 'Entertainment' is also used for obvious joke or humorous websites, such as The Onion and Uncyclopedia.
5. Forum: (websites with forum software and user accounts to post comments).

6. Governmental websites: (e.g., local, state and federal governments, usually associate with .gov, .us etc.).
7. Informational: (e.g., Wikipedia type, such as About.com, Wikipedia.org and other similar online Yellow pages).
8. News: 'News' generally indicates a certain website is part of an online news organization or a publication. (e.g., Local News site or National News - ABC, NBC, CNN, KUSI, SignonSandiego, etc.). Caveat: Some sites appear to be 'News,' but are coded as 'Special Interest Group' if it is less related to news reporting and more about postings about one specialized topic or subject.
9. NGO: (e.g., non-profit organizations, such as Red Cross, Rotary club).
10. Social media website: (e.g., such as twitter sites, Facebook, online forums), those contents are created by users directly. These are spatially accurate if the social media company is in the same location as the servers.
11. Special Interest Groups: (e.g., websites are created to promote specific concepts or items - such as political party, or issues).
12. Offline: (e.g., broken links, page not found).

The second dataset is **the geolocation verification**. Each site was visited and the mailing address of the headquarters recorded for every site. Generally, the first step was to look for the site information on the site itself first, such as on a 'Contact Us' page or in the 'Privacy Policy/Terms of Service' pages (an explicit-to-the-site address, street address is on the website). If this information could not be found, the process continued by using an outside source, such as Wikipedia (an implicit-to-the-site address, not actually on the website, but it is apparent that it is the correct street address). We used the letter [E] for explicit verification process and letter [I] for implicit verification process. The websites location scores are:

- E-1: The content creator/writer, website company, and the server (IP geolocation) are all in the same city; info found on website.
- E-2: The website company and the server are in the same city; info found on website OR the content creator/writer and the server are in the same city; info found on website.
- E-3: The website company and the server are not in the same city; info found on website.
- I-1: The content creator/writer, website company, and server are all in the same city; info found outside website.
- I-2: The website company and the server are in the same city; info found outside website OR the content creator/writer and the server are in the same city; info found outside website.

- I-3: The website company and the server are not in the same city; info found outside website.
- N/A: No geoidentifying information found, or server is at geographic center of United States, or the web page is offline.

A similar location meant that the web server and the city listed in the IP address were within **fifty miles** of each other. Fifty miles is chosen as a cutoff point that should include all entities of a particular city. These codes allowed generalized statistics on the spatial accuracy of IP address databases compared to a manual classification of each website.

## 6. Comparative Analysis of Manual Classification

Web Page Category

As discussed above, the manual classification process was used to take a more in-depth look at the CyberDiscovery results. The first part of the process was to assign a category to the web page from a domain list of twelve categories. We selected three keywords ("Mitt Romney", "HPV Vaccine", and "Flu") to illustrate the web category classification results (Figure 4).

For the "Mitt Romney" dataset the category with the highest number of records was "News", with 42.66%. The second highest category was "Blog", with 24.80%, the highest fraction for the "Blog" category of all three keywords. This is not surprising as at the time of data collection (January 4, 2012), Mitt Romney was a well-known Republican primary candidate with a high level of publicity. It would follow that a large number of news stories would be written about the candidate and that a large number of blog authors would post entries related to the favorite to win the Republican nomination.

For the "Flu" dataset, the category with the highest number of records was the "Governmental", with the second highest category being "Commercial Websites". As many of the top web pages related to the flu are part of the US CDC (United States Centers for Disease Control), the "Governmental" web pages are the top result. Regarding the high number of "Commercial Websites", there are a large number of private companies that provide cold and flu remedies, whether actually curative or not.

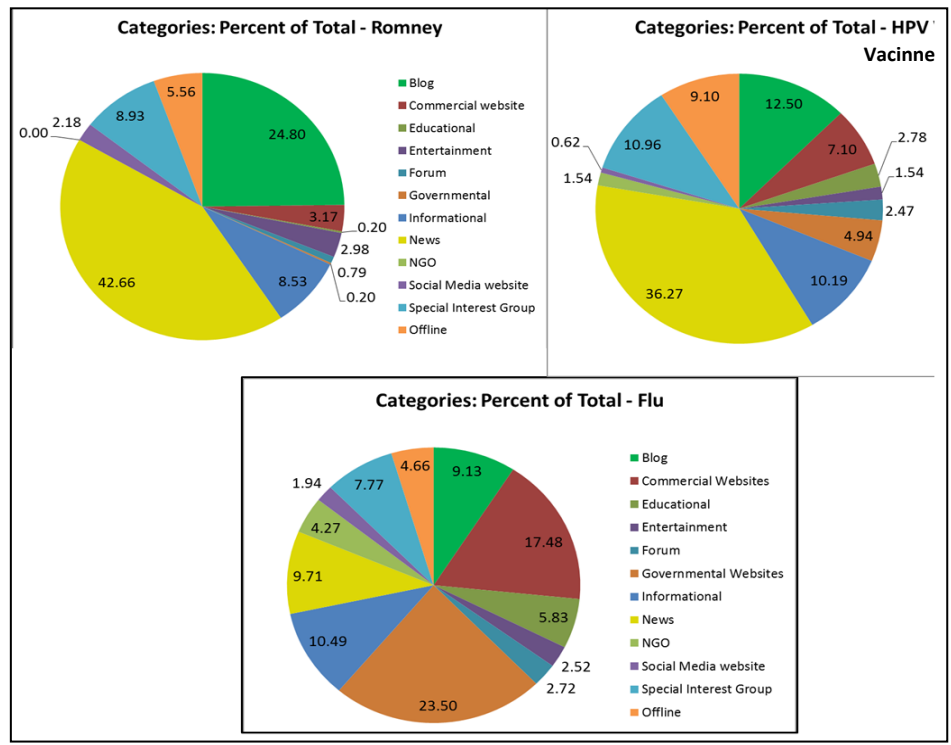

**Figure 4.** The percentages of web page categories from three sets of keyword search results ("Mitt Romney", "HPV Vaccine", and "Flu").

For the "HPV Vaccine" dataset, the category with the highest number of records was the "News" category, with "Blog" narrowly holding the second highest spot. "Special Interest Group", "Informational", and "Offline" are all close to "Blog", each containing approximately a tenth of the dataset. While it would be expected that the HPV vaccine results would be similar to the flu results, the increase in "News" web pages could be related to the fact that the HPV vaccine was very controversial when released. This controversy created a large amount of media attention that overshadowed the government affirmations of the vaccine.

Geolocational Accuracy

The third part of the process was to assign a code related to the geolocational accuracy of the web page relative to the IP address geocoded result. [E] indicates explicit verification process and [I] indicates implicit verification process. [1] means that content creator/writer, website company, and the server (IP geolocation) are all in the same city. [2] means that the website company and the server are in the same city OR the content creator/writer and the server are in the same city. [3] means that the website company and the server are not in the same city. The geolocation accuracy results can be seen in Figure 5.

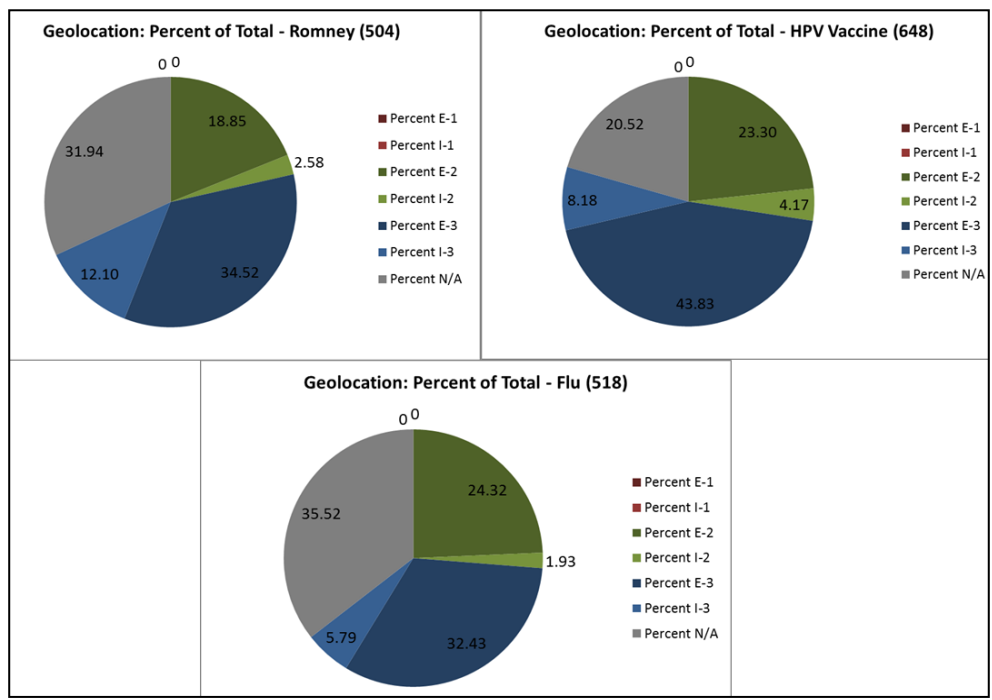

**Figure 5.** The percentages of each dataset for each code for geolocation accuracy for the results. The number in parenthesis in the title reflects the number of records in each set of results. Two types of green colors show the good geolocation results (explicit or implicit). Two blue colors show the wrong geolocation results (explicit or implicit). The gray color is the web pages without detail location information.

For the "Mitt Romney" dataset, the results clearly indicate that there is a large disconnect between the percentage of spatially accurate results and the overall dataset. Only about 20% (21.43%) of the web page IP address registrations were actually within 50 miles of the web page content creators. 46.62% were definitively in the wrong location, and 31.94% are unknown. The low spatial accuracy could be partially the result of a high number of "Blogs" and "Special Interest Groups".

For the "Flu" dataset, the spatially accurate percentage is near the others at 26.25%. This is slightly higher than "Mitt Romney". With the largest unknown class of all six datasets, at 35.52%, conclusions about the lowest spatially inaccurate class (38.22%) are not robust.

For the "HPV Vaccine" dataset, the spatially accurate percentage is in line with the other results at 27.47%. This is the second highest result, though still just over one fourth of the entire dataset. The spatially inaccurate class is large, reaching over half of the dataset at 52.01%. These results are more robust than the flu results as the unknown class is much smaller at 20.52%.

Spatial Accuracy in Different Categories of Web Pages

As would be expected, different categories have different rates of spatial accuracy. A full breakdown can be seen below in Figure 6. The percentage in Figure 6 for "Offline" is zero as all offline sites were categorized as "N/A".

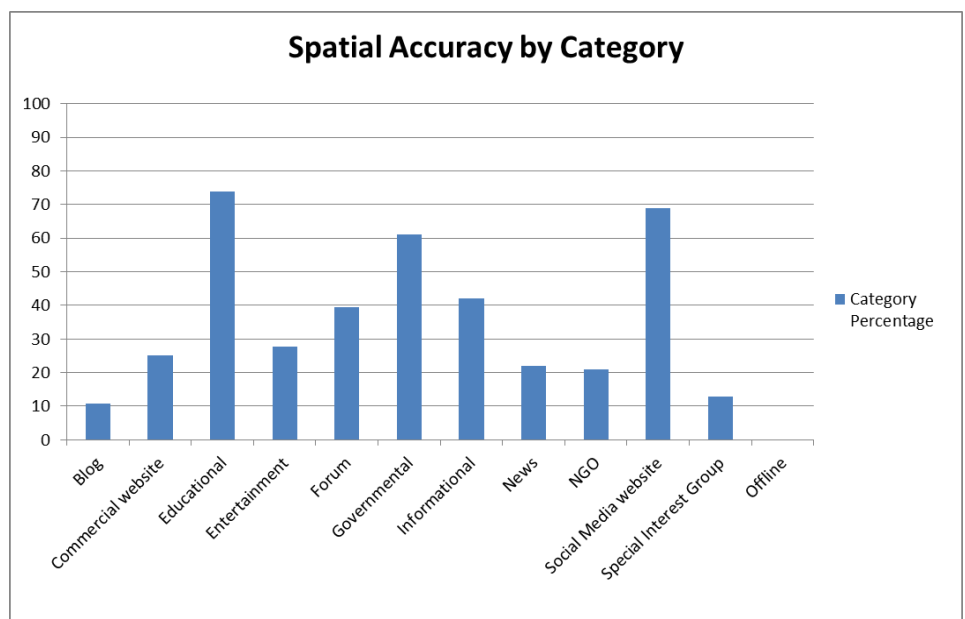

**Figure 6.** Spatial accuracy rates by category, as a percentage of all of the records in a category.

The most spatially accurate categories of web pages are "Educational" (73.86%), "Social Media website" (68.97%), and "Governmental" (60.98%). This generally makes sense, as these are categories that tend to emphasize server security over convenience. Most major universities keep server management in-house, which would lead to IP addresses registered at the university's physical address. For social media websites, most of the social media websites we picked up were major companies that manage their own servers (Facebook, Twitter, LinkedIn, etc.) as opposed to renting server space somewhere offsite. Governments would seem to be the most concerned with data integrity as they manage large amounts of very personal data (social security numbers, tax information), both at the state and local level.

The least accurate categories are "Blog" (10.81%), "Special Interest Group" (12.81%) and "NGO" (20.93%). This also is intuitive, as blogs are one of the easiest web pages to purchase. Most blogs that were gathered are hosted by WordPress or Blogspot, which are both based in the San Francisco, CA/San Jose, CA area. Since only a small percentage of blog users live in this area, only a small number of blogs can be spatially accurate. Special interest groups are similar as domain names are relatively easy to buy, allowing any group with a goal to set up a basic website that is hosted by another major company. Since many of the special interests groups gathered here were only of moderate size/influence, they did not have the need to manage their own servers. The NGO percentage is more surprising. The explanation for this may lie in methodological bias, as the delineation between "NGO" and "Special Interest Group" is relatively vague. The result may also be due to intermediate sized NGOs that are large enough to function as an NGO abroad, but are small enough that server maintenance is out of the budget.

# 7. Comparing Different Map Design Settings for Visualizing Web Information Landscapes

Several cartographic representation methods could be applied to the visualization of web page information landscapes, such as kernel density maps, choropleth maps, and graduated circle maps (Tsou et al. 2013). This study adopted the kernel density method to illustrate the density of related web pages associated with different keywords. Figure 8 illustrates the web information landscape (web page density) for three selected keyword based on our manual classification results (spatially accurate web pages) comparing to the background web page landscape. In our map design, the ranking numbers of search results were considered as the "popularity" or the "population" in the kernel density algorithm. A higher ranked web page is more "popular" and more "visible" comparing to a lower ranked web page. Therefore, we converted the ranking numbers into the population parameter for kernal density maps: **Population = (Total number of web pages + 1) - rank# (Equation 1)**. We created a "background" web information landscape in order to compare the differences between the selected keyword and the background landscape. By using 168 randomly chosen keywords and removing some stop words, we created the "background" web information landscape associated with 56,000 web pages (Tsou et al. 2013).

To calculate the differences between the keyword map and the background map, a raster-based map algebra tool from ArcGIS was used with the following formula: **Differential Value = ( Keyword-A / Maximum-Kernel-Value-of-Keyword-A ) - ( Background / Maximum-Kernel-Value-of-Background ) (Equation 2).** The web information landscape maps was created by using the manual classification results and selecting out only those records that had a score of "1" or "2" (a dataset of only spatially accurate web pages) (Figure 7). The red color indicates that the keyword web pages have higher kernel density ratio in the region comparing to the background ratio. The blue color indicates that the keyword web pages have lower kernel density ratio in the region comparing to the background ratio.

In Figure 7, some cities show higher density of web pages more than others. Cities such as Washington, DC and New York, NY tend to be higher than average. This may be due to a higher number of spatially accurate servers in these cities than average. For instance, servers from the "Government" category are more likely to be in the location assumed, as governments (federal and local) tend to maintain control over their servers as opposed to leaving them with a private company. Therefore, Washington, DC, the center of the federal government, would tend to have more servers than expected for the area. This phenomena also works in the opposite direction, as San Francisco, CA/San Jose, CA tend to be lower than average on the maps. This could be due to the large number of blogs that reside in the cities. Both Wordpress and Blogspot, two of the largest online blog communities, are based in San Francisco/San Jose. This means that IP addresses track all blogs to this area, despite the blog authors living all over the United States or the world. It is also interesting to see the hot spot of spatially accurate web pages in Utah associated with "Mitt Romney".

Since kernel density maps require some parameter configurations. We tested various combination of settings to find the best configuration for displaying web information landscapes. Different kernel sizes were tesed starting at 50,000 meters with an increment of 50,000 meters. We also compared four different population value calculation methods (zero population value, inverse rank value, logarithm of rank value, and fuzzy value). Finally, two normalization techniques were examined: the maximum score and the score range procedure.

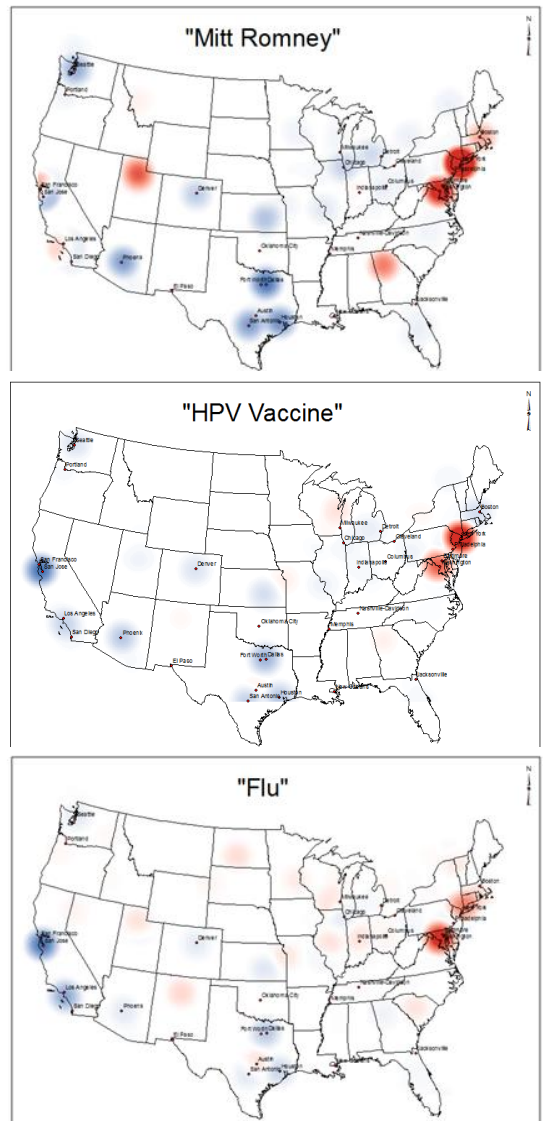

**Figure 7.** The web page information landscapes of three keywords ("Mitt Romney", "HPV Vaccine", and "Flu"). All maps created using the kernal density method with logarithmic population, a kernel radius of 200,000 meters, and the maximum score normalization. The red color indicates that the keyword web pages have higher kernel density ratio in the region comparing to the background ratio. The blue color indicates that the keyword web pages have lower kernel density ratio in the region comparing to the background ratio.

After testing various settings for kernel density maps, we recommend the following configuration for displaying web page kernal density maps: 1). The kernel radius should be between 150,000 and 200,000 meters, or a distance that does not confuse important areas on the map (in this case, nearby cities); 2). the logarithmic population method created the best results comparing to other three methods; 3). The score range procedure can create a cleaner map than the maximum score technique.

## 8. Conclusion

Within this research, we have attempted to provide a quantitative baseline for analyzing the spatial and temporal characteristics of the Yahoo search engine results. The temporal examination demonstrates a dropoff in the early results, as would be expected. However, the eventual equilibrium appears to be related to the amount of attention, or dynamism, of the search terms. If the terms are highly dynamic (such as "bin Laden"), then the eventual equilibrium is lower.

The manual classification of web pages from search engines shows that different types of keywords will generate different portions of web categories. More importantly, different categories of web pages have different spatial accuracy results. The average spatial accuracy from our datasets studied is 25.54%. These numbers are much lower than the current research estimates of IP address passive geolocation from literatures. One possible explanation is that our research focuses on the server-side IP address geolocations. Previous studies focus only on the client-side (users) IP address geolocations. There are also variations in the type of web pages and the intensity of the web pages. Higher numbers of "Educational" and "Governmental" websites and lower numbers of "Blog" and "News" web pages would increase the amount of spatially accurate results within a dataset.

While there are several options for creating visualization map from a series of points (web pages), the kernel density function has a few advantages. First, the kernel density function works well with overlapping points. Second, the mapping function works well with the other types of analysis that will be performed on the dataset, such as the population field (a kernel density function parameter) and the normalization procedure.

Overall, using IP addresses to better understand cyberspace activities is a challenging, but promising task. But with such a low spatially accuracy rate for using IP addresses (ranging from 25% to 35%) to geolocate web pages, we need to carefully examine and explain these web information landscapes (kernel density maps). Some further research are needed to answer a few key questions. For example, are the 25%-35% spatially accurate points (web pages) clustered or randomly distributed on maps? Will a kernel density map reveal the spatial hot spots of the 25% aaccurate points along with the 75% noisy data? Are the rest of 75% error points (web page) randomly distributed or clusted on a map? These research questions are very common in conducting Big Data analytics and web data mining tasks. Many Big Data sources (such as web search engines, social media, and crowd sourcing data) can not get 100% accurate datasets. We need to use data mining tools and data filtering procedures to remove the errors and noises from datasets effectively. Then we can apply various visualization methods to discover spatial and temporal patterns of Big Data, including web pages, social media, and volunteered geographic information.

## Acknowledgement

This work is supported by the Center for Human Dynamics in the Mobile Age (HDMA), San Diego State Universirty and the U.S. National Science Foundation under Grant No. 1028177 and Grant No.1416509, project titled, "Spatiotemporal Modeling of Human Dynamics Across Social Media and Social Networks". Any opinions, findings, and conclusions or recommendations expressed in this material are those of the author(s) and do not necessarily reflect the views of the National Science Foundation.